\def\yboxit#1#2{\vbox{\hrule height #1 \hbox{\vrule width #1
\vbox{#2}\vrule width #1 }\hrule height #1 }}
\def\fillbox#1{\hbox to #1{\vbox to #1{\vfil}\hfil}}
\def\ybox{{\lower 1.3pt \yboxit{0.4pt}{\fillbox{8pt}}\hskip-0.2pt}}

\def\l{\left}
\def\r{\right}
\def\comments#1{}

\def\half{{1\over 2}}

\def\tr{{\rm tr\ }}

\def\rrangle{\rangle\rangle}
\def\llangle{\langle\langle}

\def\CO{{\cal O}}

\def\II{\relax{I\kern-.10em I}}

\def\IZ{\relax\ifmmode\mathchoice
{\hbox{\cmss Z\kern-.4em Z}}{\hbox{\cmss Z\kern-.4em Z}}
{\lower.9pt\hbox{\cmsss Z\kern-.4em Z}}
{\lower1.2pt\hbox{\cmsss Z\kern-.4em Z}}\else{\cmss Z\kern-.4em
Z}\fi}
\def\IB{\relax{\rm I\kern-.18em B}}
\def\IC{{\relax\hbox{$\inbar\kern-.3em{\rm C}$}}}
\def\ID{\relax{\rm I\kern-.18em D}}
\def\IE{\relax{\rm I\kern-.18em E}}
\def\IF{\relax{\rm I\kern-.18em F}}
\def\IG{\relax\hbox{$\inbar\kern-.3em{\rm G}$}}
\def\IGa{\relax\hbox{${\rm I}\kern-.18em\Gamma$}}
\def\IH{\relax{\rm I\kern-.18em H}}
\def\II{\relax{\rm I\kern-.18em I}}
\def\IK{\relax{\rm I\kern-.18em K}}
\def\IP{\relax{\rm I\kern-.18em P}}

\def\inbar{\,\vrule height1.5ex width.4pt depth0pt}

\font\cmss=cmss10 \font\cmsss=cmss10 at 7pt
\def\IR{\relax{\rm I\kern-.18em R}}

\def\rk{{r}}
\def\chch{\hbox{ch}}

\def\aroof{\hat A}
\def\lroof{\hat L}

\def\pocl{\hbox{p}}

\def\BR{\IR}
\def\BZ{Z} 
\def\BP{\IP}
\def\BR{\IR}
\def\BC{\IC}

\def\lp10{l_P^{10}}
\def\lp11{l_P^{11}}
\def\R11{R_{11}}

\documentstyle[12pt]{article}

\begin{document}

\input FEYNMAN

\begin{titlepage}
\begin{flushright}
\parbox{3.5cm}{CITUSC/01-044\\
hep-th/0111086\\
November 2001}
\end{flushright}

\vspace{2pt}
\begin{center}
{\Large \bf (Fractional) Intersection Numbers, Tadpoles and Anomalies}

\vspace{1cm}

{Christian R\"omelsberger}

\vspace{5mm}

{\em Department of Physics and Astronomy\\
and\\
CIT-USC Center for Theoretical Physics\\
University of Southern California\\
Los Angeles, CA 90089-0484, USA}
\end{center}

\vspace{5pt}
\centerline{{\bf{Abstract}}}
\vspace{5pt}
\begin{small}
We use the Witten index in the open string sector to determine tadpole 
charges of orientifold planes and D-branes. As specific examples we 
consider type I compactifications on Calabi Yau manifolds and noncompact 
orbifolds. The tadpole constraints suggest that the standard embedding 
is not a natural choice for the gauge bundle. Rather there should be 
a close connection of the gauge bundle and the spin bundle. In the 
case of a four fold, the standard 
embedding does not in general fulfill the tadpole conditions. We show 
that this agrees with the Green-Schwarz mechanism. In the case of 
noncompact orbifolds we are able to solve the tadpole constraints 
with a gauge bundle, which is related to the spin bundle. We compare 
these results to anomaly cancellation on the fixed plane of the 
orbifold. In the case of branes wrapping noncompact cycles, there are 
fractional intersection numbers and anomaly coefficients, which we 
explain in geometric terms.
\end{small} 
\end{titlepage}

\section{Introduction}

There has been a lot of progress in understanding D-branes in type~II sting 
theory compactified on Calabi-Yau manifolds, even away from the geometrical
regime. A first step in this direction was to identify charges of D-branes 
in nongeometrical regimes with geometrical charges (see e.g. 
\cite{bdlr,dg,dr,kllw,scheid}). In subsequent work the dynamics of these 
D-branes was understood further, (see e.g. 
\cite{dfra,dfrb,diadoug,mayr,govind,tomasiello,mikecath}). All these 
approaches are dealing with D-branes which are point particles with 
different charges in transverse space.

In this note we want to put forward some foundations for the use of the 
previously described methods, in the context of space filling branes in 
type~I theory. One of the differences is that these cannot be inserted in 
arbitrary numbers, but they have to fulfill some constraints due to 
the inconsistent RR flux they can produce. The cancellation of such 
tadpoles has been considered in numerous papers (see e.g. 
\cite{grscb,gipo,gijo,blin,abiu,afiv,urang}).

The intersection of two D-branes is characterized by a massless fermionic
string stretching between the two branes. Similarly, in the case of a 
D-brane and an O-plane the intersection is characterized by a massless 
fermionic string stretching between the D-brane and its image. The
chirality of these fermionic strings determines the sign of the
intersection. This intersection product is actually the intersection
product of quantum K-theory. 

We use this observation to determine the nontorsion tadpole charges 
of D-branes and O-planes, by calculating the Witten index of open 
string interactions with various 'probe' branes in a low energy 
description of the branes. This leads to a very quick procedure to determine
the tadpole charges. 

In geometric Calabi Yau compactifications as
well as in orbifold compactifications the tadpole constraints suggest
a natural connection between the gauge bundle and the Dirac spinor bundle
on the compactification manifold. This is opposed to the usual solution
of the tadpole constraints on a 3-fold in terms of the standard
embedding. In the case of $\BC^d/\BZ_N$ orbifolds we find an explicit 
expression for the gauge bundle in terms of the spin bundle.

For noncompact spaces there arises another surprise. The intersection 
numbers between D-branes wrapping noncompact cycles turn out to be 
fractional. This is due to the continuous spectrum of momentum states in 
the noncompact directions. Geometrically this can be understood as coming 
from torsion of these cycles at the boundary of the noncompact space.

In section \ref{chapinters} we derive the general expressions for 
intersection numbers between two D-branes and between a D-brane and an
O-plane. In section \ref{chapgeom} we apply these general results to
geometric Calabi-Yau compactifications and find agreement with 
anomaly cancellation. Finally in section \ref{chaporb} we explore 
orbifold compactifications. We find fractional intersection numbers
between noncompact cycles and explain these in terms of relative homology.
We solve the tadpole constraints for some noncompact orbifolds and
compare these results to anomaly cancellation on the orbifold fixed
plane.

\section{Intersection numbers and tadpole charges}
\label{chapinters}

To illustrate the connection between intersection numbers and charges,
let us look at type IIB string theory compactified on a two torus $T^2$.
In order to determine in which way a D-string is winding around the
torus, it is enough to determine its intersection numbers with two
known D-strings wrapping the two fundamental 1-cycles of the torus.

At each intersection point of two D-strings on the torus, there is a 
massless fermionic string located. The chirality of this string 
indicates the sign of the intersection. In the following, we will 
generalize this concept of intersection numbers to more complicated
brane configurations and also to orientifold planes. 

In order to do this we want to describe a general formalism to calculate
the torsion free part of the quantum K-theory charge of an orientifold
plane. This can be done using the Witten index in the open string sector
to define an intersection number \cite{bedo,df,bdlr}.

In the CFT, the orientifold plane is described by a crosscap state and
D-branes are described by boundary states, the closed string states which 
are emitted. The RR charge of a D-brane/O-plane is given by the 
RR ground state part of the corresponding boundary/crosscap state. 
To determine the RR charge it is useful to define a nondegenerate 
intersection form between boundary/crosscap states.

Given two boundary states $|E\rrangle$ and $|F\rrangle$ the intersection
is defined as the overlap of the RR ground state part of the two
boundary states, with a $(-)^{F_R}$ inserted in order to make it topological:
\begin{eqnarray}
I(E,F)&=&\langle E, RR-gs|(-)^{F_R}|F, RR-gs\rangle=\nonumber \\
&=&\llangle E, RR|(-)^{F_R} e^{-2\pi t H^{(cl)}} |F, RR\rrangle.
\end{eqnarray}
This (topological) cylinder amplitude can also be calculated in the open 
string sector
\begin{equation}
\label{intersb}
I(E,F)=\tr_{R_{E,F}}(-)^F e^{-{\pi\over t} H^{(op)}_{E,F}}.
\end{equation}
This is in general easier than the calculation in the closed string
sector and can often be done in the low energy theory of the D-branes.

Similarly, the intersection number between a crosscap state  $|C\rrangle$ 
and a boundary state $|E\rrangle$ is the M\"obius amplitude
\begin{equation}
I(C,E)=\llangle C, RR|(-)^{F_R} e^{-2\pi t (L_0-{c\over 24})}
e^{-2\pi t (\tilde L_0-{c\over 24})} |E, RR\rrangle.
\end{equation}
Doing a modular transformation to the open string sector gives
\begin{equation}
\label{intersd}
I(C,E)=\tr_{R_{E^*,E}}\Omega (-)^F e^{-{\pi\over 4t} (L_0-{c\over 24})}.
\end{equation}
The world-sheet parity operator $\Omega$ has to satisfy $\Omega^2=1$ in this
open string sector and the Hilbert space can be divided into positive and 
negative parity eigenspaces. 

\section{The geometric Calabi Yau compactification}
\label{chapgeom}

\subsection{Tadpole analysis}
\label{geotadpole}

To show how the formalism that we explained above works, we repeat the 
calculation of \cite{scruccaa} for geometric Calabi Yau compactifications. 
In the geometric case D-branes can be thought of as Chan-Paton
bundles on the $2d$ dimensional Calabi Yau space $X$. The open 
string Ramond ground states between two D-branes $E$ and $F$ can be 
described by harmonic sections of 
\begin{equation}
\Delta\otimes E^* \otimes F,
\end{equation}
where $\Delta$ is the Dirac spinor bundle over $X$. The fermion number
operator $(-)^F$ acts as the chirality operator on the spinors. From this 
can see that the intersection number (\ref{intersb}) is the index of
the twisted Dirac operator:
\begin{equation}
I(E,F)=\int_X \chch(E^*\otimes F) \aroof(R).
\end{equation}
This is the K-theoretic intersection number \cite{mimo}.

The action of $\Omega$ on open strings exchanges Chan-Paton factors in 
the fundamental representation of the gauge group on the one end of the 
string with Chan-Paton factors in the antifundamental representation of 
the gauge group on the other end of the string, i.e. it acts on the ends 
of an open string by $E\mapsto E^*$. This means that the open string Ramond 
ground states in the M\"obius amplitude are harmonic sections of 
$\Delta\otimes E\otimes E$. $\Omega$ acts on these states simply by 
transposition $\gamma\mapsto\gamma^t$. In this way $\Delta\otimes E\otimes E$ 
is divided into the two Eigenspaces of $\Omega$ with Eigenvalues~$\pm 1$
\begin{equation}
\Delta\otimes E\otimes E=\Delta\otimes S^2E\oplus\Delta\otimes\Lambda^2E.
\end{equation}

The M\"obius amplitude is now
\begin{equation}
Z_M=\int_X \chch(S^2E)\aroof(R)-\int_X \chch(\Lambda^2E)\aroof(R).
\end{equation}
To fully calculate this expression we have to relate the Chern characters
in the symmetric and antisymmetric representation to the Chern Characters
of the fundamental representation
\begin{eqnarray}
\chch(S^2E)&=&\half(\chch^2(E)+\chch(2E)),\nonumber \\
\chch(\Lambda^2E)&=&\half(\chch^2(E)-\chch(2E)),
\end{eqnarray}
where $\chch(2E)$ means that the curvatures in the expression for
the Chern character are multiplied by $2$. This gives the M\"obius 
amplitude
\begin{equation}
Z_M=\int_X \chch(2E)\aroof(R)=2^d\int_X \chch(E)\aroof\l({R\over 2}\r).
\end{equation}
Using trigonometric theorems the M\"obius amplitude can be expressed as
\begin{equation}
Z_M=2^d\int_X \chch(E)\sqrt{\aroof(R)}\sqrt{\lroof\l({R\over 4}\r)},
\end{equation}
which shows that the Mukai charge of the crosscap state is 
$2^d\sqrt{\lroof\l({R\over 4}\r)}$. For the full string theory $2^d$ can 
actually be replaced with $2^d\times 2^{5-d}=32$ where the $2^{5-d}$ comes
from the transverse dimensions.

In order to cancel the tadpoles one has to introduce a boundary state
with a $Spin(32)/\BZ_2$ gauge bundle $E$ satisfying 
$\chch(E)\sqrt{\aroof(R)}=32\sqrt{\lroof\l({R\over 4}\r)}$, i.e.
\begin{eqnarray}
\label{bundlecharges}
\pocl_1(E)&=&\pocl_1(T), \nonumber \\
\pocl_2(E)&=&-{1\over 8}\pocl_2(T)+{15\over 32}\pocl_1^2(T)
\end{eqnarray}
in cohomology. These conditions might be trivially satisfied if $X$ has 
low enough dimension. For example, the 8-form condition only applies on 
a 4-fold.

There is another interesting way to look at these conditions. Using the
splitting principle the Chern character of the bundle $E$ can be expressed
as
\begin{equation}
\chch(E)=2^d\prod_j\cosh{x_j\over 4},
\end{equation}
where the Dirac spinor bundle of $X$ has the Chern character
\begin{equation}
\chch(\Delta)=2^d\prod_j\cosh{x_j\over 2}.
\end{equation}
This suggests, that the natural way to build the gauge bundle $E$ is actually
related to the spin bundle $\Delta$ and not via the standard embedding, which
in general does not fulfill tadpole cancellation on a 4-fold. We will see a 
very similar condition later in the context of noncompact orbifolds. There 
we will be able to find an explicit solution to the analogous condition.

\subsection{Comparison to the Green-Schwarz mechanism}

The result from tadpole cancellation looks a little bit surprising from
the point of view of the Green-Schwarz mechanism \cite{grsc}. But as we
will see in this section there arises the same 8-form condition from
the Green-Schwarz mechanism.
In the 10 dimensional type I supergravity there are three contributions
to the chiral anomaly, the gravitino, the dilatino and the gauginos.
Apart from these chiral fermions, there is a 2-form field $B$ with a
3-form field strength $H$, which is used to cancel the chiral anomaly.
The anomaly can be canceled, if the anomaly polynomial factorizes into
a 4-form and a 8-form part
\begin{equation}
\hat I_{12}=X_4\wedge X_8.
\end{equation}
giving rise to the Bianchi identity
\begin{equation}
dH=X_4
\end{equation}
and the equation of motion
\begin{equation}
d\ast H=X_8.
\end{equation}

These two equations imply that in the absence of 1- and 5-branes,  
the integral of $X_4$ and $X_8$ around any compact cycle has to 
vanish. It is not surprising that this agrees with the 4-form and 
8-form conditions (\ref{bundlecharges}) from tadpole cancellation 
(see also equation. (3.46) in \cite{freed}).

In the context of heterotic strings it looks like the standard embedding 
should always work. This is only a statement in the perturbative nonlinear
sigma model description. There should be inconsistencies appearing in the
presence of NS5-branes \cite{urang,duff}. NS5-branes are the magnetic duals 
to the F-string and the 8-form condition is a condition on spacefilling F-strings.

\section{Noncompact orbifolds}
\label{chaporb}

Let us now apply the same ideas to noncompact orbifolds $\BC^d/\Gamma$ with 
isolated singularities only. The gauge theory part of such string theories 
can be described in terms of quiver diagrams \cite{dm,dgm}.

The new feature here is that we are dealing with space filling D$(2d)$-branes, 
which are not localized on the orbifold singularities and by that token 
are not described by a four dimensional effective theory. The orbifold
group has an action on $\BC^d$ as well as on the Chan-Paton factors. 
The different irreducible representations on the Chan-Paton factors are
as usual denoted by vertices of a quiver diagram.
The arrows on the other hand behave slightly different than in the 
case of localized fractional branes. This is due to the fact that the
open strings stretching between two such branes are not localized at
the singularity and can propagate in the noncompact directions. This
momentum part has nontrivial transformation properties under the orbifold
group and can make up for some nontrivial transformation properties of
the zero mode part.

In addition to this outer quiver, there is, of course, the well known inner 
quiver, describing fractional D0-branes. The fractional D0-branes represent
branes wrapped around compact cycles \cite{dm,ddg}.

\subsection{Fractional intersection numbers in $\BC^d/\Gamma$}
\label{orbtadpcharge}

To calculate the number of arrows between two fractional D$(2d)$-branes one 
can make use of the character valued index theorem \cite{goodm,egh}. For 
simplicity let us take a $\BC^d/\BZ_N$ orbifold, where $\BZ_N$ acts on 
$\BC^d$ in a diagonal way 
$(z_1,\cdots,z_d)\mapsto(\epsilon_1 z_1,\cdots,\epsilon_d z_d)$, 
where $\epsilon_j^N=1$ and $\prod\epsilon_j=1$. For each irreducible 
representation of $\gamma$ there exists a type of fractional D$(2d)$-brane, 
which has this irreducible representation acting on its Chan-Paton factors. 
In the case of a $\BZ_N$ orbifold these are the multiplication with $N$ 
different phases.

To calculate the intersection numbers between different fractional branes, 
it is sufficient to keep only the massless open fermionic strings
which propagate in $\BC^d/\Gamma$. They are characterized by their Chan-Paton
factors $\mu$ and $\nu$, spinor degrees of freedom and a momentum. The 
Witten index (\ref{intersd}) is now a trace over
these massless fermionic strings with the chirality operator $\Gamma^{2d+1}$
inserted. In flat space this trace vanishes, because the trace over
the spinors vanishes, and it is surprising that in the
untwisted sector of an orbifold theory this is not true. This comes about 
because the orbifold group action, which has to be inserted into the trace
contains gamma matrices.

The index consists of three different traces, the trace over the spinors,
the trace over the Chan-Paton factors and the integral over moment in the
noncompact directions. For each complex 
direction, the action of the orbifold group on the spinors is given by
\begin{equation}
\label{gammaaction}
\cos{\alpha_j\over 2}+\sin{\alpha_j\over 2}\Gamma_{2j}\Gamma_{2j+1},
\end{equation}
where $\epsilon_j=e^{i\alpha_j}$. In order that the trace over the spinors is
nonvanishing the second term in (\ref{gammaaction}) has to be picked up. This 
gives a contribution of $(-i)^d2^d\prod\sin{\alpha_j\over 2}$ from the trace 
over the spinors. The next contribution comes from the momentum integral
\begin{equation}
\int dp^{2d}\delta^{(2d)}(gp-p)={1\over|\det(1-g)|}=
{1\over 4^d\prod\limits_j\sin^2{\alpha_j\over 2}}.
\end{equation}
Finally there is a contribution  $e^{2\pi i{\mu-\nu\over N}}$ from the 
trace over the Chan-Paton factors. Putting all this together and summing 
over the orbifold group gives:
\begin{equation}
\label{orbinters}
I^{(o)}_{\mu\nu}={(-i)^d\over N}\sum_m{}^\prime{e^{2\pi i{\mu-\nu\over N}}\over 
2^d\prod\limits_j\sin{\alpha_j m\over 2}},
\end{equation}
where the prime indicates that the sum is only over terms which have a 
nonvanishing denominator, i.e. terms where the trace over the spinors does 
not vanish.

This expression for the intersection numbers is hard to simplify, but it 
will turn out in the following that they are fractional. The 
nonintegrality of the index itself is not inconsistent, because there is 
no energy gap between the ground states and states with momentum in the 
noncompact orbifold directions \cite{semenoff}. In the cases where the 
noncompact orbifold $\BC^d/\Gamma$ can be embedded into a compact 
orbifold $T^{2d}/\Gamma$, the intersection numbers between D$(2d)$-branes 
on the noncompact orbifold can be calculated by dividing the (integer) 
intersection number on the compact orbifold by the number of fixed 
points. These results agree with (\ref{orbinters}).

In order to understand these fractional intersection numbers better, it is
useful to calculate the intersection numbers of the inner quiver.
They can be calculated in a similar manner. The only difference is the 
absence of the momentum integral:
\begin{equation}
\label{orbinterss}
I^{(i)}_{\mu\nu}={(-i)^d\over N}\sum_m{e^{2\pi i{\mu-\nu\over N}}
2^d\prod\limits_j\sin{\alpha_j m\over 2}}.
\end{equation}
By expanding the $\sin$'s in terms of exponential functions it is easy to
see that this is the same result as from counting the invariant chiral
fields in the quiver gauge theory (see e.g. \cite{dm,dgm}).

The intersection forms $I^{(i)}$ and $I^{(o)}$ are both degenerate and 
have a null vector corresponding to the pure D0-brane and the pure 
D$(2d)$-brane (regular representation). Taking the quotient of the 
two intersection `lattices' by the null vectors, it is easy to see 
that the two lattices are, up to a sign  of $(-)^d$, inverse to each 
other. This explains the fractionality of $I^{(o)}$. 

The full intersection matrix of both, the inner and the outer quiver is
\begin{equation}
\label{complorbinters}
I=\left(\begin{array}{cc} I^{(o)} & 1 \\ (-)^d & I^{(i)} \end{array}\right).
\end{equation}
The rank of $I$ is $N+1$. This shows, that there are only $N+1$ independent
charges, either the pure D0-charge and N D$(2d)$-charges or the other way around.

\subsection{Geometric explanation}

The fractionality of the intersection numbers seems from the geometric point
of view a little bit surprising, but it can be understood quite naturally
in terms of relative homology. To illustrate the basic idea, it is useful to 
consider the example of $\BC^2/\BZ_2$. The blow up of this orbifold is the
total space of the line bundle $\CO(-2)\stackrel{\pi}{\longrightarrow}\BP^1$.

There are two compact cycles, the point and the $\BP^1$. It is easy to see,
that the point doesn't intersect with any other compact cycle, but the $\BP^1$
has a self intersection $-2$. This can be seen from the zeroes of a section
of the normal bundle $\CO(-2)$. The intersection matrix for the compact
cycles is then
\begin{equation}
I^{(c)}=\left(\begin{array}{cc} 0 & 0 \\ 0 & -2 \end{array}\right).
\end{equation}

The fractional D$(2d)$-branes are described by Chan-Paton bundles over
the noncompact space $\CO(-2)$. The bundles of interest are pull backs
of line bundles over the base $\BP^1$
\begin{equation}
\begin{array}{c}
\pi^*\CO_{\BP^1}, \\
\pi^*\CO_{\BP^1}(1).
\end{array}
\end{equation}
These branes have lower charges, which are the pull back of a point
onto the fiber, i.e. the fiber itself. It is useful to keep track of the
behavior at infinity of such a noncompact 2-cycle. The boundary at 
infinity of $\CO(-2)$ is $S^3/\BZ_2=\BR P^3$ and the boundary of a
fiber of $\CO(-2)$ is a noncontractible torsion 1-cycle in $\BR P^3$.
A 2-cycle which is wrapping the fiber twice has a trivial boundary 
in $\BR P^3$ and can be contracted to a compact 2-cycle, a $\BP^1$.

From the argument above it is easy to see, that the intersection 
number of two noncompact 2-cycles wrapping the fiber of $\CO(-2)$ is
${-2\over2\cdot 2}=-\half$. This is the inverse of the nonzero 
eigenvalue of $I^{(c)}$.

We can now generalize this argument. Let $X$ be
a noncompact Calabi-Yau manifold and let $\dot X$ be it's boundary
at infinity. Then compact cycles are described by the ordinary
homology $H_*(X,\BZ)$. In order to describe noncompact cycles, it is
useful to keep track of the behavior at infinity. This is done by the
relative homology $H_*(X,\dot X,\BZ)$.

An element of $H_*(X,\dot X,\BZ)$ is denoted by the equivalence class
$[\Gamma,\gamma]$, where $\Gamma\subset X$ and $\gamma\subset\dot X$. The 
relative boundary operator acts on a chain as
\begin{equation}
(\Gamma,\gamma)\mapsto\partial(\Gamma,\gamma)=
(\partial\Gamma-\gamma,-\partial\gamma).
\end{equation}
This can be understood as subtracting the boundary of $\Gamma$ inside $\dot X$ 
from the regular boundary of $\Gamma$. The condition for a cycle to be closed,
implies that $\partial\Gamma=\gamma$ and $\partial\gamma=0$ as expected.
The equivalence relation then becomes
\begin{equation}
[\Gamma,\gamma]\sim[\Gamma+\partial\Lambda-\lambda,\gamma-\partial\lambda].
\end{equation}
It is easy to see that $\Lambda$ is the usual homology equivalence and 
$\lambda$ is some piece in the boundary $\dot X$ that can be added.

It is easy to see that there is an exact sequence
\begin{equation}
\begin{array}{ccccc}
H_p(X,\BZ) & \stackrel{i}{\longrightarrow} & H_p(X,\dot X,\BZ) & 
\stackrel{r}{\longrightarrow} & H_{p-1}(\dot X,\BZ) \\
{[\Gamma]} & \mapsto & {[\Gamma,0]} & & \\
& & {[\Gamma,\gamma]} & \mapsto & {[\gamma]}
\end{array}
\end{equation}
Any cycle $[\Gamma,\gamma]$ which restricts to torsion on the boundary, can
be multiplied by the order $N$ of the torsion\footnote{Note that even though
$[\gamma]$ is a torsion element in $H_*(\dot X,\BZ)$, $[\Gamma,\gamma]$ is
not necessarily a torsion element in $H_*(X,\dot X,\BZ)$.}. 
$[N\Gamma,N\gamma]$ is then an element in $H_*(X,\BZ)$ and intersection 
numbers are well defined. Fractions are produced due to the multiplication 
by $N$.

Since cycles in $H_*(X,\BZ)$ are only in the interior of $X$, there is also a 
natural, integral intersection product between elements of $H_*(X,\BZ)$ 
and $H_*(X,\dot X,\BZ)$. This intersection product is nondegenerate 
\cite{spanier}. The intersection lattice $H_*(X,\BZ)$ modulo the null
vectors is the dual to the intersection lattice of elements in 
$H_*(X,\dot X,\BZ)$ which restrict to torsion on the boundary. There 
is an analogous statement in cohomology \cite{guvawi,botttu}, which 
is equivalent by Poincare duality.

This explanation also works quite well for $\BC^3/\BZ_3$, which has 
$S^3/\BZ_3$ as boundary. The homology on $S^3/\BZ_3$ has been worked 
out \cite{gurawi}, it has \linebreak[4] $H_1(S^3/\BZ_3,\BZ)=\BZ_3$ and 
$H_3(S^3/\BZ_3,\BZ)=\BZ_3$. The resolved orbifold is the total space 
of the bundle $\CO(-3)\stackrel{\pi}{\longrightarrow}\BP^2$
and the arguments are very similar to the ones in $\BC^2/\BZ_2$.

\subsection{Solution to the tadpole constraints}

In order to measure the charge of a crosscap state one can either calculate
it's intersection with D0-probes or D$(2d)$-probes. For comparison with anomaly 
cancellation on the fractional D9-branes it is useful to consider D$(2d)$-probes.

In order to calculate the intersection product of a probe brane with a 
crosscap state we first have to determine the action of $\Omega$ on 
the R ground states. The action can be divided into three parts, the 
trivial action on the momentum, the action on the Chan-Paton factors 
and the action on the spinor indices.

The action on the fermion zero modes is $\psi_0^m\mapsto\pm\psi_0^m$, 
depending on whether there are N-N or D-D boundary conditions in the 
$m$-th direction. If there is an even number of D-D directions, then 
$\Omega$ acts as a chirality operator in these directions \cite{gipo}. 
For the case of D$(2d)$-branes this means that the $\Omega$ action on the
spinors is trivial.

The requirement that the intersection of a D$(2d)$-brane probe $\mu$ with the 
crosscap state is the same as its intersection with the tadpole canceling 
D$(2d)$-brane boundary state can be summarized as
\begin{equation}
\label{tadpolecond}
I_{\mu\Omega(\mu)}^{(o)}=\sum_\nu w_\nu I_{\nu\mu}^{(o)},
\end{equation}
together with the requirement of having a total of $2^d$ fractional 
D$(2d)$-branes. Because we omitted the $10-2d$ transverse dimensions, there
is a factor of $2^{5-d}$ missing on the right hand side of the equation.
The final result has to be multiplied by $2^{(5-d)}$.

One would expect that for a high enough rank of the orbifold group, 
the equations (\ref{tadpolecond}) might not have a solution with 
nonnegative integer numbers $w_\nu$ of fractional D$(2d)$-branes, leading to 
inconsistent backgrounds for type I theory. In the examples we consider, 
this actually turns out not to be the case.

A solution to these equations is given by a $2^d$ dimensional Chan-Paton 
representation of the orbifold group. The decomposition of this representation 
into irreps specifies the multiplicities of fractional D$(2d)$-branes. The 
dimension of such a representation suggests a close connection
to (Dirac) spinor representations.

Indeed for the $\BC^d/\BZ_n$ orbifolds this guess turns out to be true. 
Equation (\ref{tadpolecond}) can be written as 
\begin{equation}
{(-i)^d\over N} \sum_m{}^\prime e^{2\pi i{2\mu\over N} m} 
{1\over 2^d\prod\limits_j\sin{\alpha_j m\over 2}}=
\sum_\nu w_\nu {(-i)^d\over N} \sum_m{}^\prime e^{2\pi i{\mu-\nu\over N} m}
{1\over 2^d\prod\limits_j\sin{\alpha_j m\over 2}}
\end{equation}
Using that $N$ is odd and discrete Fourier transformation, this expression
can be converted to 
\begin{equation}
\sum_\nu w_\nu e^{2\pi i{2\nu\over N} m}=
2^d\prod\limits_j\cos{\alpha_j m\over 2}.
\end{equation}
The left hand side of this equation is a sum over the characters of all
irreps $2\nu$ of $\BZ_N$ with multiplicities $w_\nu$. The right hand side
is the character of the Dirac spinor representation associated to the 
geometrical action of $\BZ_N$. This shows that the Chan-Paton representation
is almost the Spinor representation, except for a `reshuffling' of
the characters on the left hand side. This situation is similar to what we 
have seen in the geometric case in section \ref{geotadpole}.

\subsection{Local anomalies in orbifolds}

In order to check anomaly cancellation in orbifold theories, we need 
to derive an expression for the anomaly on a fixed plane of an orbifold 
due to the untwisted 
fields of the theory. An anomaly on the fixed plane due to fields from 
the twisted sector can be calculated in a straight forward way by the 
descent formalism. The derivation in this section is very similar to 
the derivation of the index in section~\ref{orbtadpcharge}, but we want 
to do this derivation in a bit more detail, because it is also quite 
important for a more elementary understanding of anomalies in orbifolds 
of M-theory \cite{howia,howib,wiorb}.

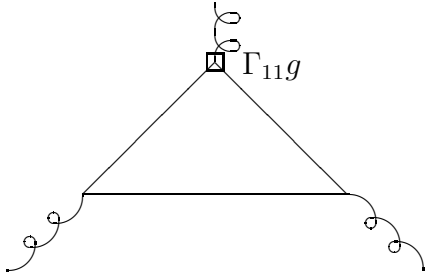
\begin{figure}
\centerline{
\begin{picture}(16000,11000)
\drawline\fermion[\E\REG](3000,3000)[10000]
\drawline\fermion[\NE\REG](3000,3000)[7070]
\drawline\fermion[\NW\REG](13000,3000)[7070]
\drawline\gluon[\N\REG](\fermionbackx,\fermionbacky)[2]
\drawline\gluon[\SW\REG](3000,3000)[2]
\drawline\gluon[\SE\REG](13000,3000)[2]
\put(7700,7700){\framebox(600,600){}}
\put(8700,7400){\makebox(2900,900){$\Gamma_{11}g$}}
\end{picture}
}
\caption{The anomalous diagram for an orbifold fixed plane.}
\label{anomdiagram}
\end{figure}

Typically the one loop chiral anomaly due to ten dimensional fields is 
calculated with the help of a hexagon diagram \cite{alvgomwi}, giving 
rise to a ten dimensional anomaly. The anomaly on a $(10-2d)$-dimensional 
orbifold fixed plane has to be derived from a $(6-d)$-polygon diagram with 
the ten dimensional chiral fields running around in the loop. This can be 
nonvanishing because of the insertion of the gamma matrices from the orbifold 
group action $g$ in the loop (see figure~\ref{anomdiagram}). The momentum 
integral is still over a ten dimensional momentum.

The momenta and polarizations of the external lines can be set in the 
direction of the orbifold fixed plane, then the traces in the diagram 
can be split into traces inside the fixed plane and traces transverse 
to the fixed plane. The traces transverse to the fixed plane are actually 
the same traces that lead to the index (\ref{orbinters}). The traces 
inside the orbifold fixed plane are exactly the same as for the chiral 
anomaly of a $(10-2d)$-dimensional fermion. So we conclude that the chiral 
anomaly on the fixed plane from the ten dimensional fermion fields is the 
chiral anomaly of a $(10-2d)$-dimensional fermion with the index 
(\ref{orbinters}) as a prefactor.

To check this result and to understand the fractionality of the prefactor 
(\ref{orbinters}), we want to look at some elementary examples. These 
are all orbifolds of tori. The $T^6/\BZ_3$ for example has 27 fixed planes. 
The anomaly on a single fixed plane is equal to the 
four dimensional chiral anomaly of the dimensionally reduced theory. This 
four dimensional anomaly is equally distributed over all 27 fixed
planes. In the dimensionally reduced theory there are three chiral fermions 
between two different fractional D9-branes, which leads to a prefactor of 
${3\over 27}={1\over 9}$ for the chiral anomaly of each fixed plane. This 
agrees with the result from (\ref{orbinters}). One can look at the orbifolds 
$T^4/\BZ_2$ and $T^4/\BZ_3$ in a similar way.

The chiral anomaly of the gravitino can be calculated in a similar way. 
There are two different contributions from the gravitino, one where the 
vector index of the gravitino is in the orbifold fixed plane, it is 
invariant under the orbifold group, this gives rise to a 
$(10-2d)$-dimensional gravitino anomaly with a prefactor 
(\ref{orbinters}). If the vector index is transverse to the orbifold 
fixed plane, the orbifold group acts on it with the regular 
representation. This case is treated in the same way as spin $\half$ 
fermions.

A ten dimensional gravitino gives in this way rise to $(10-2d)$-dimensional 
'gravitino' and 'dilatino' anomalies. In the case of 
$d=3$ it is easy to see that the four dimensional anomalies due to the ten 
dimensional gravitino vanish, because the sum (\ref{orbinters}) is over an 
odd function.

To conclude this section we want to make a quick comment on chiral anomalies
in orbifolds of M-theory \cite{howia,howib,wiorb}. For M-theory on 
$\BR/\BZ_2$ the relevant diagram is the hexagon diagram. The orbifold 
group acts on spinors with a $\Gamma_{11}$, and a prefactor of $\half$ 
comes from the rank of the orbifold group. This shows that on each 
fixed plane in the Horava-Witten picture there is half of the ten 
dimensional anomaly. A similar argument applies for M-theory on $T^5/\BZ_2$.

\subsection{Anomalies in quiver theories}

Now we want to investigate the anomaly cancellation in the case of
noncompact orbifolds. In the case of compact orbifolds there have been 
quite detailed studies (see e.g. \cite{blpssw}).
The anomaly conditions in the case of quiver theories are a bit
weaker than in the ten dimensional case. We consider a theory which
is compactified on an orbifold down to four dimensions.

Away from the orbifold singularities, anomaly cancellation implies
the same as for smooth compactifications.
The anomaly polynomial on a four dimensional fixed plane is a 6-form 
polynomial of the form
\begin{equation}
\hat I_6=I[\chch(E_{tot})\aroof(R)]_6,
\end{equation}
$I$ being the prefactor (\ref{orbinters}).
Each term in this polynomial has to factorize into 2-form and 4-form 
parts. The term which potentially might not factorize this way is
$\tr F^3$ for any factor of the gauge group. Only unitary factors of
the gauge group actually have a nonvanishing $\tr F^3$. If such a 
$U(N)$ factor has $N>2$ then this trace (in the fundamental 
representation) does not factorize and creates an anomaly which cannot
be canceled by a closed string twist field living on the
fixed plane.

In the orientifolded theory such $U(N)$ factors arise from vertices
which are not fixed under the $\BZ_2$ involution $\Omega$. Such 
a vertex $\mu$ can have two different kinds of arrows ending on it, 
arrows which are fixed under the involution and arrows which are 
not fixed. The latter ones give rise to matter in the fundamental or 
anti-fundamental representation of the gauge group $U(w_{\mu})$ 
depending on the direction of the arrow.

The fixed arrows give rise to antisymmetric representations. 
The Chern characters 
in the antisymmetric representations can be expressed in terms 
of Chern characters in the fundamental or antifundamental representation of 
$U(w_\mu)$ depending again on the direction of the arrow.
\begin{eqnarray}
\chch_3(\Lambda^2E)&=&\rk(E)\chch_3(E)-4\chch_3(E)+\cdots, \nonumber \\
\chch_3(\Lambda^2E^*)&=&-\rk(E)\chch_3(E)+4\chch_3(E)+\cdots,
\end{eqnarray}
where the dots indicate terms which don't involve $\chch_3(E)$.

The constant in front of $\chch_3(E)$ in the anomaly polynomial
can be written as
\begin{equation}
\label{quanomalyb}
\sum_{\nu\ne\Omega(\mu)} w_\nu I^{(o)}_{\nu\mu}+w_\mu I^{(o)}_{\mu\Omega(\mu)}
- 4 I^{(o)}_{\mu\Omega(\mu)}=
\sum_\nu w_\nu I^{(o)}_{\nu\mu}- 4 I^{(o)}_{\mu\Omega(\mu)}.
\end{equation}
The vanishing of
(\ref{quanomalyb}) follows from tadpole cancellation (\ref{tadpolecond}), 
but the tadpole condition is stronger, because there are also conditions
for fixed vertices and $U(N\le 2)$ gauge groups.

\section*{Acknowledgments}
I would like to thank Per Berglund, Andy Brandhuber, Ilka Brunner, 
Rich Corrado, Mirjam Cvetic, Emanuel Diaconescu, Mike Douglas, 
Jaume Gomis, Nick Halmagy, 
Greg Moore and Nick Warner for helpful discussions. I'm grateful to 
the Rutgers Physics Department where this project started.
This work was supported in part by funds provided by the DOE under
the grant number DE-FG03-84ER-40168.

\pagebreak

\bibliography{tadpole}
\bibliographystyle{unsrt}

\end{document}